\documentclass[12pt, reqno]{amsart}
\usepackage[foot]{amsaddr}
\usepackage{times}
\usepackage[super,numbers,sort&compress]{natbib}
\makeatletter
\renewcommand{\email}[2][]{%
  \ifx\emails\@empty\relax\else{\g@addto@macro\emails{,\space}}\fi%
  \@ifnotempty{#1}{\g@addto@macro\emails{\textrm{(#1)}\space}}%
  \g@addto@macro\emails{#2}%
}
\makeatother

\usepackage[margin=1in]{geometry}
\usepackage[utf8]{inputenc}
\usepackage{tikz}
\usepackage{natbib}
\usepackage{amsmath}
\usepackage{amssymb}
\usepackage{amsfonts}

\begin{document}
\title{Economies and Diseconomies of Scale in Segmented Mobility Sharing Markets}
\author{Hongmou Zhang}
\author{Xiaotong Guo}
\author{Jinhua Zhao}
\address{\textup{H.Z. is with the School of Government, Peking University, Beijing 100871, China, and was with the Singapore--MIT Alliance for Research and Technology Centre, Singapore 138602.  X.G. is with the Department of Civil and Environmental Engineering, Massachusetts Institute of Technology, MA, USA 02139. J.Z. is with the Department of Urban Studies and Planning, Massachusetts Institute of Technology, MA, USA 02139. (Corresponding author: Xiaotong Guo.)}}
\email{zhanghongmou@pku.edu.cn, \{xtguo, jinhua\}@mit.edu}
\begin{abstract}
 On-demand mobility sharing, provided by one or several transportation network companies (TNCs), is realized by real-time optimization algorithms to connect trips among tens of thousands of drivers and fellow passengers. In a market of mobility sharing comprised of TNCs, there are two competing principles, the economies of network scale, which can be maximally achieved with monopoly, and the healthy competition between TNCs, which requires more than one TNC in the market, but which can also lead to ``segmentation'' of market. To understand the substantiality and relationship of the two competing principles, we need to answer how much efficiency loss is generated due to the segmentation of market, and which factors are related to it. Here we show how four critical factors of market structure and characteristics of mobility sharing services---density of trips (thickness), maximum detour allowed for sharing (tightness), market shares (unevenness), and spatial segregation of the TNCs (dissolvedness)---are associated with the efficiency loss, represented as the difference in vehicle miles traveled (VMT) under different market structures. We found that 1) while VMT shows a simple power function with thickness, the corresponding exponent term can be expressed as a non-monotonic function with tightness---essentially showing how economies and diseconomies of scale in this market arise, and appearing a very similar form to the Lennard--Jones model in inter-molecular potentials \cite{Jones1924}; and 2) the efficiency loss is higher when unevenness is closer to 0.5 (50-50 market share) and dissolvedness is larger. Our results give a comprehensive analysis of how the inefficiency of market segmentation is generated, and how potentially it may be avoided through market mechanism design. Our model uses mobility sharing market as a specific example, but the competing effect between the economies and diseconomies of scale is a very general phenomenon in economics, and we anticipate our model to be able to illustrate a potential mechanical explanation in general.
\end{abstract}
\maketitle

On-demand mobility sharing is a transportation mode that matches travelers, including both drivers and passengers, for their trips by one or several transportation network companies (TNCs), such as Uber, Lyft, Didi, Via, and Grab, using real-time matching optimization algorithms \cite{Santi2014}. In mobility sharing, the use of network effect \cite{metcalfe2013metcalfe}---the more people use a TNC platform, the more likely that each of them will be matched for their trips---to generate economies of scale is well acknowledged. The economies of network scale of matching can save the number of cars needed \cite{Alonso-Mora2017,Kondor2022}, save parking space \cite{Kondor2019}, and reduce vehicle miles traveled (VMT) in a city \cite{Santi2014}. However, as in other markets, where monopoly efficiency is also studied, the lack of competition can also lead to problems, including price discrimination \cite{varian1985price}, abuse of dominance \cite{gal2013abuse}, and the lack of equity consideration \cite{Ge2016,Edelman2017}. Therefore, there exists a natural conflict between the two principles---the economies of network scale and the healthy competition between TNCs.

However, there is still no well established model about how much economies-of-scale efficiency in a \emph{segmented} market, where there are more than one TNC, is lost due to the missing opportunities of trips which should be matched but could not be because being on different platforms. A better understanding of the explicit form of this trade-off could better help us evaluate the cost and benefit of monopoly in this market. Thus, we would like to answer the following two questions: 1) how the two forces are competing with each other in generating the efficiency and inefficiency of the market of mobility sharing; and 2) what factors may change this efficiency/inefficiency.

In this paper, we for the first time, give a comprehensive closed-form model regarding the inefficiency generated due to the segmentation of mobility sharing market, and use the model to explicitly explain how the economies of scale, and the diseconomies of scale are dragging the inefficiency in opposite directions. First, we used the taxi data of Manhattan to build a mobility sharing network \cite{Santi2014}, and then simulated the fellow passenger matching of the identical demand but in two markets: 1) a monopolistic market, and 2) a duopolistic market. Second, by comparing the total VMTs in the two markets, a proxy for efficiency, the explicit amount of inefficiency is calculated. Third, we model and explain the relationship of the inefficiency with four market structure and sharing service parameters: the trip density, the detour allowed for sharing, the market shares of the two TNCs, and the spatial segregatedness of the two platforms in terms of service areas. We denote the four parameters as four ``-nesses,'' as described in Table \ref{tab:notation}.
\begin{table}[h!]
\caption{Notations for major variables used in this paper}
\centering
\footnotesize
 \begin{tabular}{c | c | l } 
 \hline
 Notation & Term & Physical Explanation \\
 \hline 
 VMT$_0$& No-sharing VMT& Vehicle miles travelled without sharing\\
  VMT$_1$& Monopoly VMT&Vehicle miles travelled with one TNC\\
   VMT$_2$& Duopoly VMT &Vehicle miles travelled with two non-cooperative TNCs\\
 $\nu$ & Thickness & Trip density as a share (from 0 to 1)\\
 $\delta$& Tightness & Delay time cap of sharing (seconds)\\ 
 $\sigma$ & Unevenness &Market shares of the duopoly TNCs  \\
 $\rho$ & Dissolvedness &Spatial dissolvedness of trips of the two TNCs \\
   $\ell$ & Efficiency loss & Efficiency loss between duopoly market and monopoly market ($\mbox{VMT}_2 - \mbox{VMT}_1$)\\
 \hline
 
 \end{tabular}
\label{tab:notation}
\end{table}

\section{Economies of Scale with Trip Thickness}
\begin{figure}[!ht]
\centering
\includegraphics[width=3in, trim=0.3in 0.3in 0.5in 0.5in, clip]{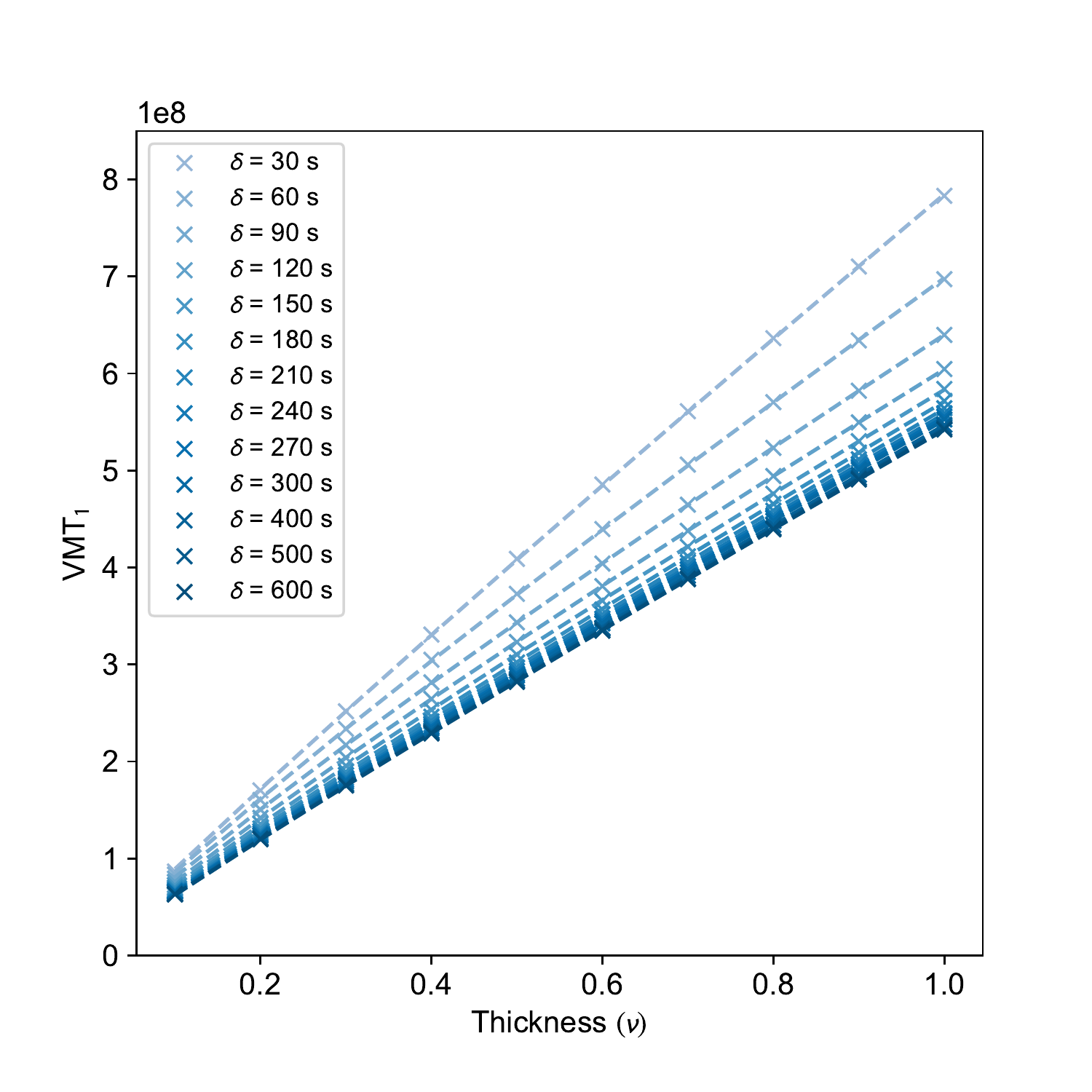}
\put(-215,210){\footnotesize\fontfamily{phv}\selectfont (a)}
\includegraphics[width=3in, trim=0.3in 0.3in 0.5in 0.5in, clip]{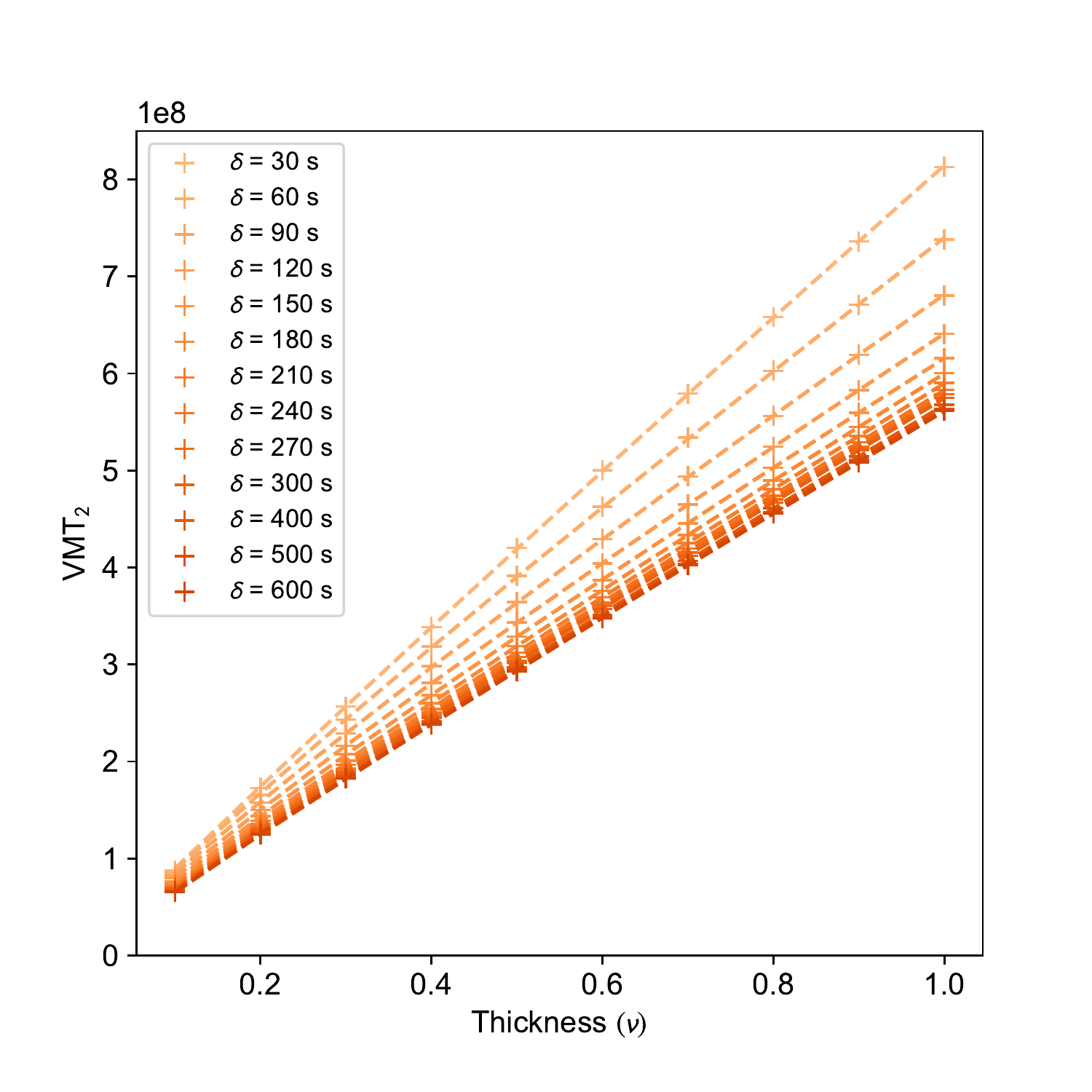}
\put(-215,210){\footnotesize\fontfamily{phv}\selectfont (b)}

\includegraphics[width=3in, trim=0.1in 0 0.5in 0.3in, clip]{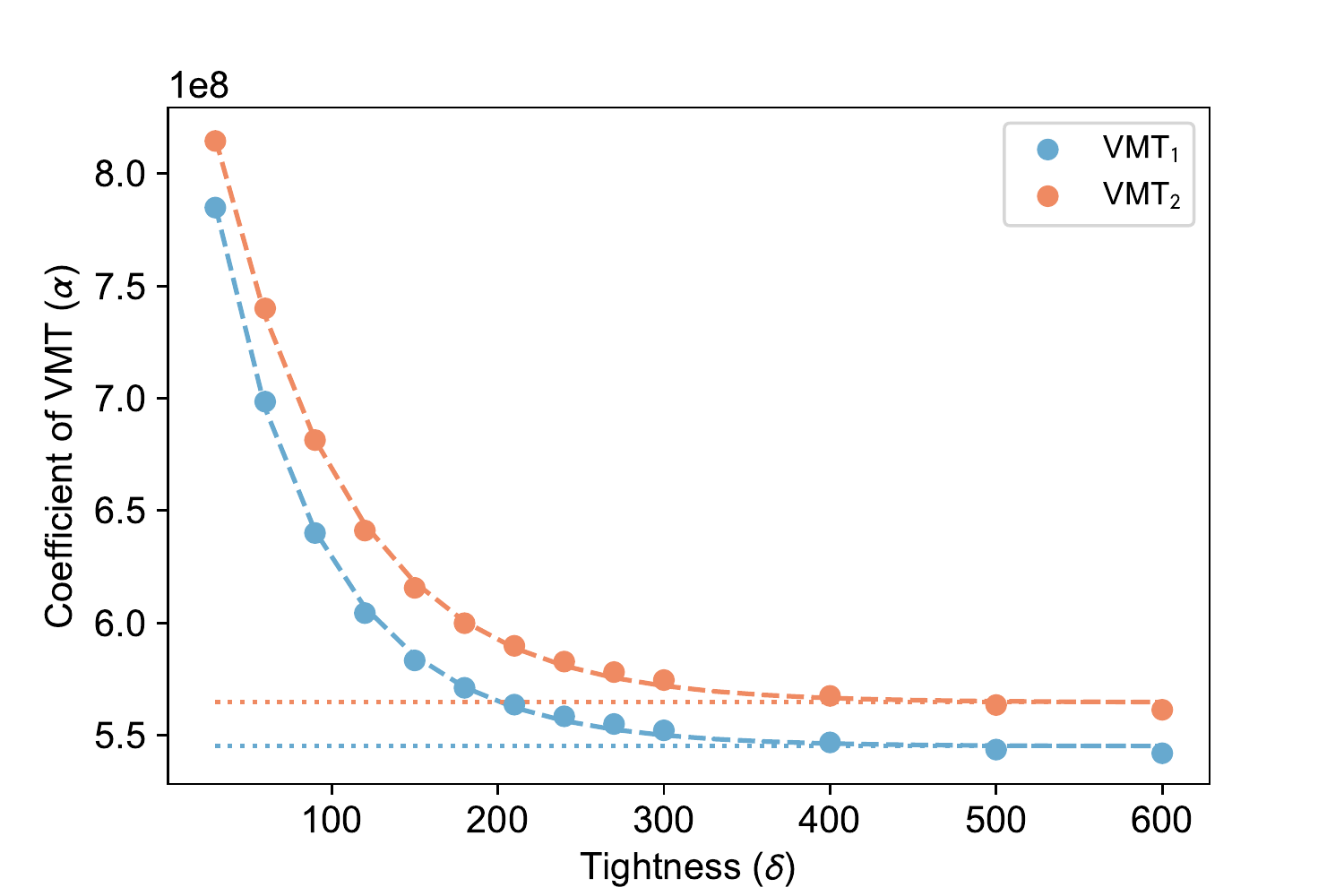}
\put(-215,145){\footnotesize\fontfamily{phv}\selectfont (c)}
\includegraphics[width=3in, trim=0.1in 0 0.5in 0.3in, clip]{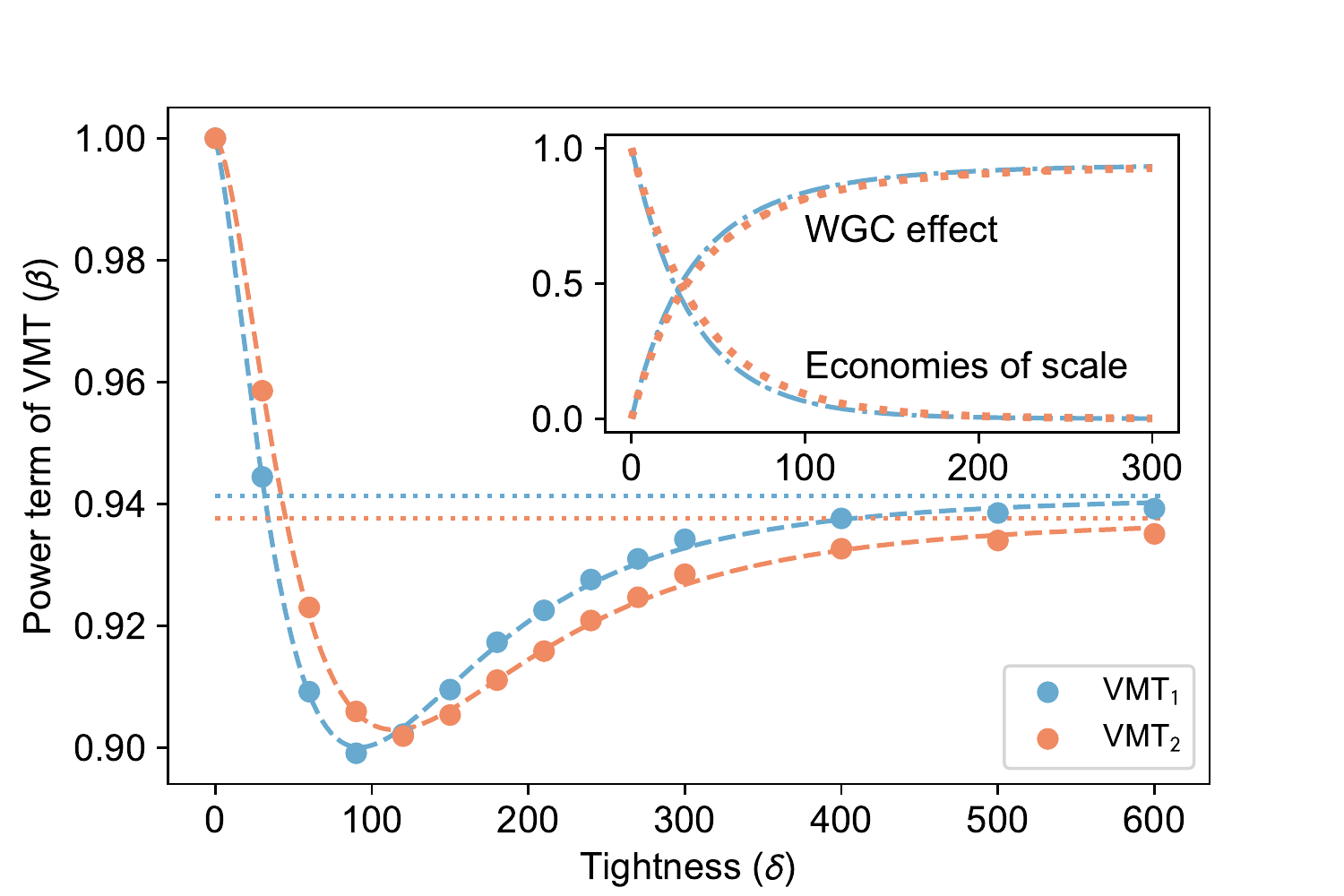}
\put(-215,145){\footnotesize\fontfamily{phv}\selectfont (d)}

\includegraphics[width=3in, trim=0.3in 0.2in 0.5in 0.5in, clip]{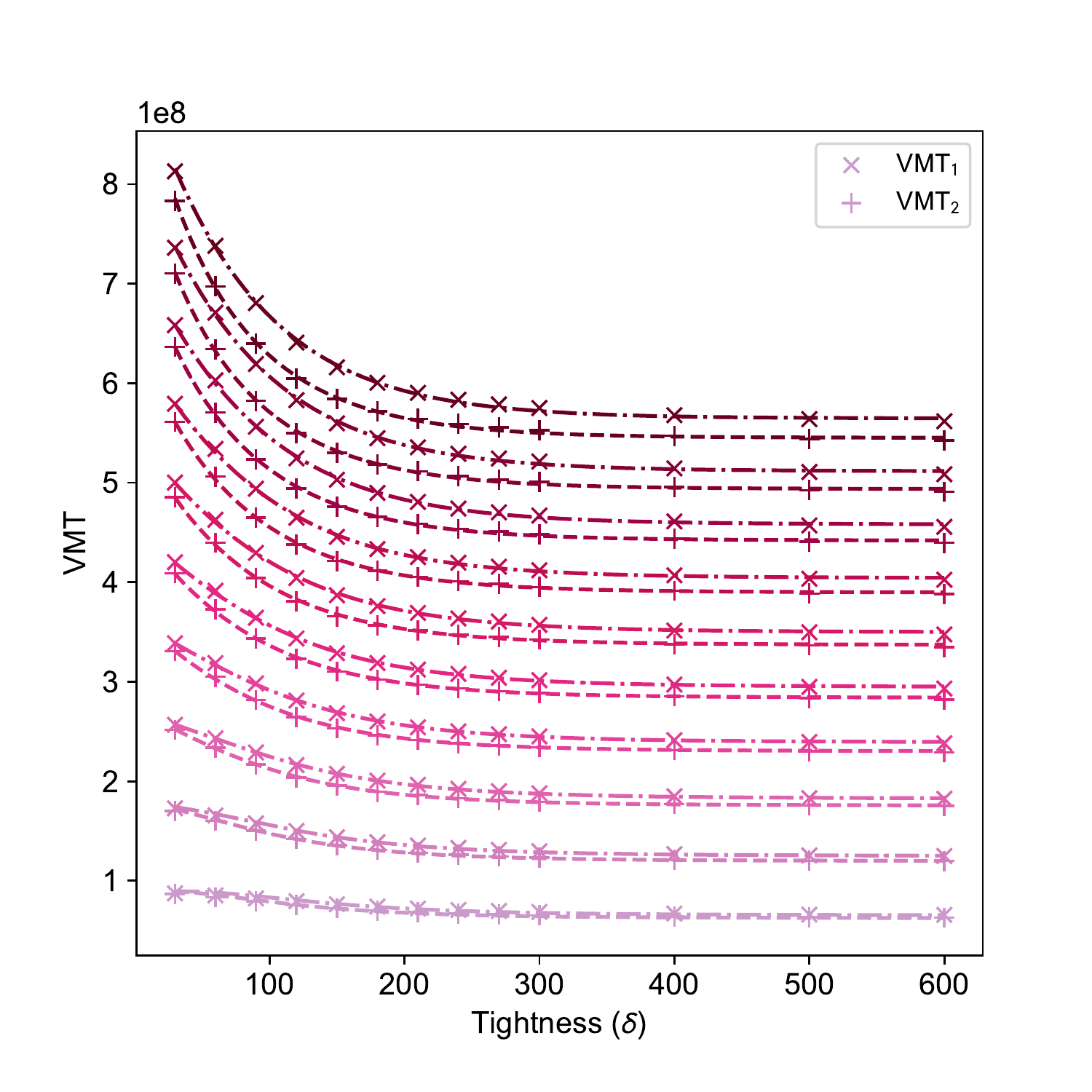}
\put(-215,220){\footnotesize\fontfamily{phv}\selectfont (e)}
\includegraphics[width=3in, trim=0.3in 0.2in 0.5in 0.5in, clip]{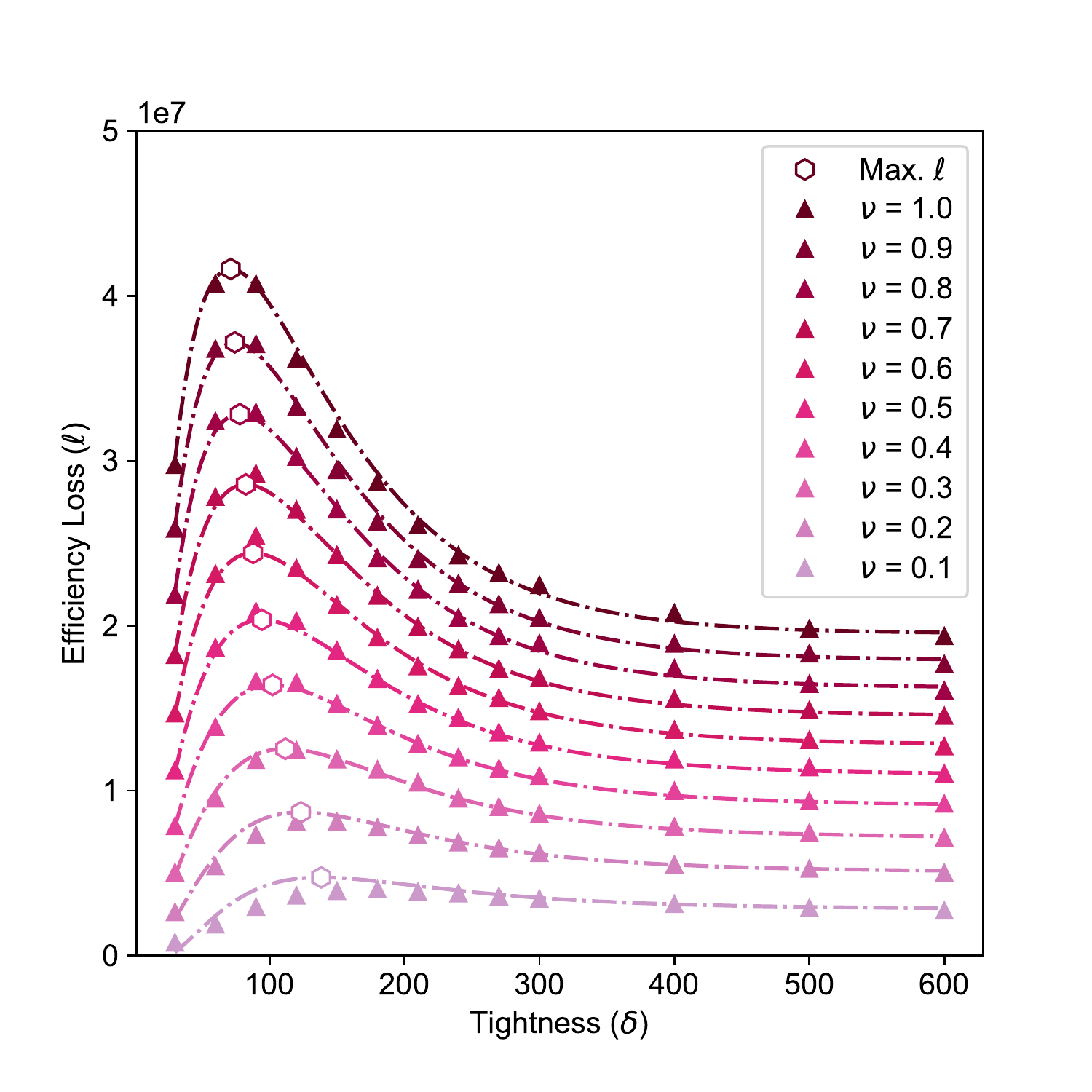}
\put(-215,220){\footnotesize\fontfamily{phv}\selectfont (f)}
\caption{Results (add caption description)}
\label{fig:mainresults}
\end{figure}

As a first-order attempt, to explore how the economies of scale are manifested, we plotted the relationship between VMTs in both the monopoly (VMT$_1$) and the duopoly market (VMT$_2$) with the density of trips ($\nu$), sampled from 0 to 100\% of the trips in the Manhattan dataset. The relationships are shown in Fig. \ref{fig:mainresults}a-b. The different curves in each graph correspond to different maximum detours allowed, i.e., tightness ($\delta$). While it appears that the relationship is linear, especially for small $\delta$'s, the curvature starts to manifest while the value of $\delta$ increases. The concave curves indicate that while the trip density increases, the increase of VMT becomes slower. This is exactly the economies of scale--induced efficiency in mobility sharing markets---when more people join the market, the matching rate increases, and the marginal VMT increase slows down. Moreover, the economies of scales are more pronounced while larger maximum detours are allowed ($\delta$), since it further expands the possibility of more successful matchings.

To quantify this economies-of-scale--induced curvature, we used the exponential functions (Eq. \ref{eq:vmt-nu}), which also show superior goodness-of-fit compared with linear models (see: Methods). Two coefficients $\alpha$ and $\beta$ are used for each curve fixing a $\delta$. 
\begin{equation}
    \mbox{VMT} = \alpha \nu^{\beta}
    \label{eq:vmt-nu}
\end{equation}
\
\section{Two Competing Forces Manifested}
After we obtained the $\alpha$'s and $\beta$'s, the  economies-of-scale metrics for all $\delta$'s, a natural extension inquiry arises---how the metrics are connected with the corresponding $\delta$. Therefore, we further treated these two coefficients as functions of $\delta$, and plotted the relationship for each, as shown in Fig. \ref{fig:mainresults}c-d. While it is straightforward that $\alpha (\delta)$ indicates the relative ``heights'' of the curves in Fig. \ref{fig:mainresults}a-b, and the curves converge as $\delta$ is sufficiently large---almost all trips are already matched above such detour flexibility (and $\alpha(\delta)$ is well fitted with a negative-exponential function, see Eq. \ref{eq:alpha-delta}, where $a_i$'s are coefficients), the curves of $\beta (\delta)$ are nontrivially non-monotonic.
\begin{equation}
    \alpha(\delta) = \frac{a_1}{e^{a_2\delta}+a_3} + a_4
    \label{eq:alpha-delta}
\end{equation}

As discussed above, the connotation of $\beta$ is the level of  economies of scale given a $\delta$. Thus, the fall and rise of the $\beta (\delta)$ curve indicates that the economies of scale decrease and increase with the tightness of market---the maximum detour allowed. It appears to be a combined effect of two curves, which drives our direction of inquiry. After trials of fitting the curves with well-known functions, we found that the Lennard--Jones functions \cite{Jones1924} representing intermolecular potentials\footnote{For two molecules which are far from each other, the force between them is primarily attractive, and the attractive force becomes stronger when they move closer; but when they are too close from each other, i.e., within a certain threshold, the force becomes repulsive. } surprisingly fit well (Eq. \ref{eq:beta-delta}), albeit with different exponent values. The mathematical decomposition of the first two terms also gives a quick analogy of economic interpretation---the curve is a \emph{bona fide} combination of an ``attractive'' force and a ``repulsive'' force. The falling part (the first term in the equation) represents the attractive force of economies of scale---when there are more matching flexibilities allowed (longer detours allowed), the matching rate also increases, and the economies of scale are more pronounced.
\begin{equation}
    \beta(\delta) = \frac{1}{(b_1 \delta+1)^m} - \frac{b_2}{(b_3\delta+1)^n} +b_2
    \label{eq:beta-delta}
\end{equation}

On the reverse side, after a certain threshold of tightness is passed (in the Manhattan dataset, roughly 100 seconds), the economies of scale are counteracted by the repulsive force. In the mobility sharing case, this effect has been denoted as the ``wild good chase''\cite{Castillo2017}---when sufficient supply is provided, the marginal cost of providing additional service increases because then they will be assigned to distant demand, i.e., the wild geese. The physical interpretation is that when more flexibilities are allowed, nearby passengers can be easily matched, but \emph{passengers who are far from each other will also more matched}. Therefore, as more detours are introduced, they eventually canceled some VMT savings due to matchings of distant passengers, showing a negative effect on the economies of scale.
\begin{figure}[!ht]
    \centering
    \includegraphics[width=3in]{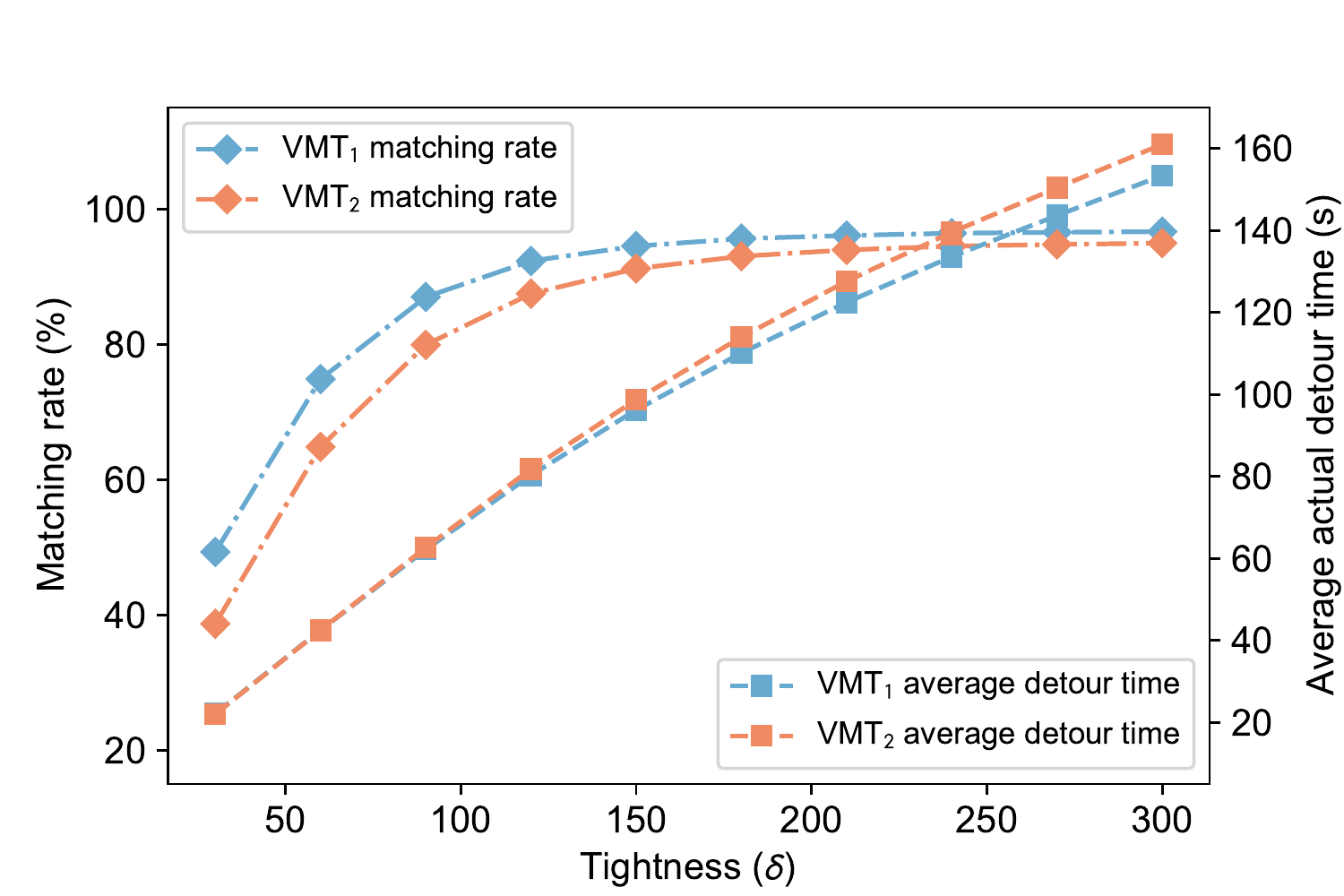}
    \caption{Matching rates and average actual detour times}
    \label{fig:twoforces}
\end{figure}
To prove this formal decomposition to be truly explainable to what really happened in the mobility sharing market, we examined the change of matching rate and detour time with tightness, shown in Fig. \ref{fig:twoforces}, and the observation confirms our intuition above---the matching rate increases and converges, and the detour keeps increasing, together shaping the combined curve---behaving exactly like the attractive and the repulsive forces between molecules.

The above findings are for the monopoly and duopoly market individually. What will happen if we consider the difference between them, the inefficiency caused by market segmentation? To understand the relative inefficiency, we put the VMTs in both markets in the same graph, Fig. \ref{fig:mainresults}e. Since we have discussed the non-linearity of the VMT with regard to tightness and thickness in each market, it is very unlikely that the VMT difference will be linear to either of the two, which is also confirmed in the graph. However, Given the absolute value of VMT to be large, the inefficiencies only show as small slivers between each pair of corresponding curves in the two markets, though not trivial. Therefore, instead of looking at it at the scale of total VMTs, we plot the relationship of efficiency loss, denoted as $\ell(\nu,\delta) = \mbox{VMT}_2-\mbox{VMT}_1$, with $\delta$ and $\nu$ in Fig. \ref{fig:mainresults}f, which further manifests the non-linearity and non-monotonicity. To examine where the maximum efficiency loss occurs, 
we may let the derivatives $\partial\ell/\partial \delta =0$ to get the closed-from positions, showing as the hexagons in the graph (see detailed forms in the Supplementary Information).
\section{Unevenness and Dissolvedness}
\begin{figure}[!ht]
\centering
    \includegraphics[height=2.1in,trim=0.1in 0.2in 0.55in 0.5in, clip]{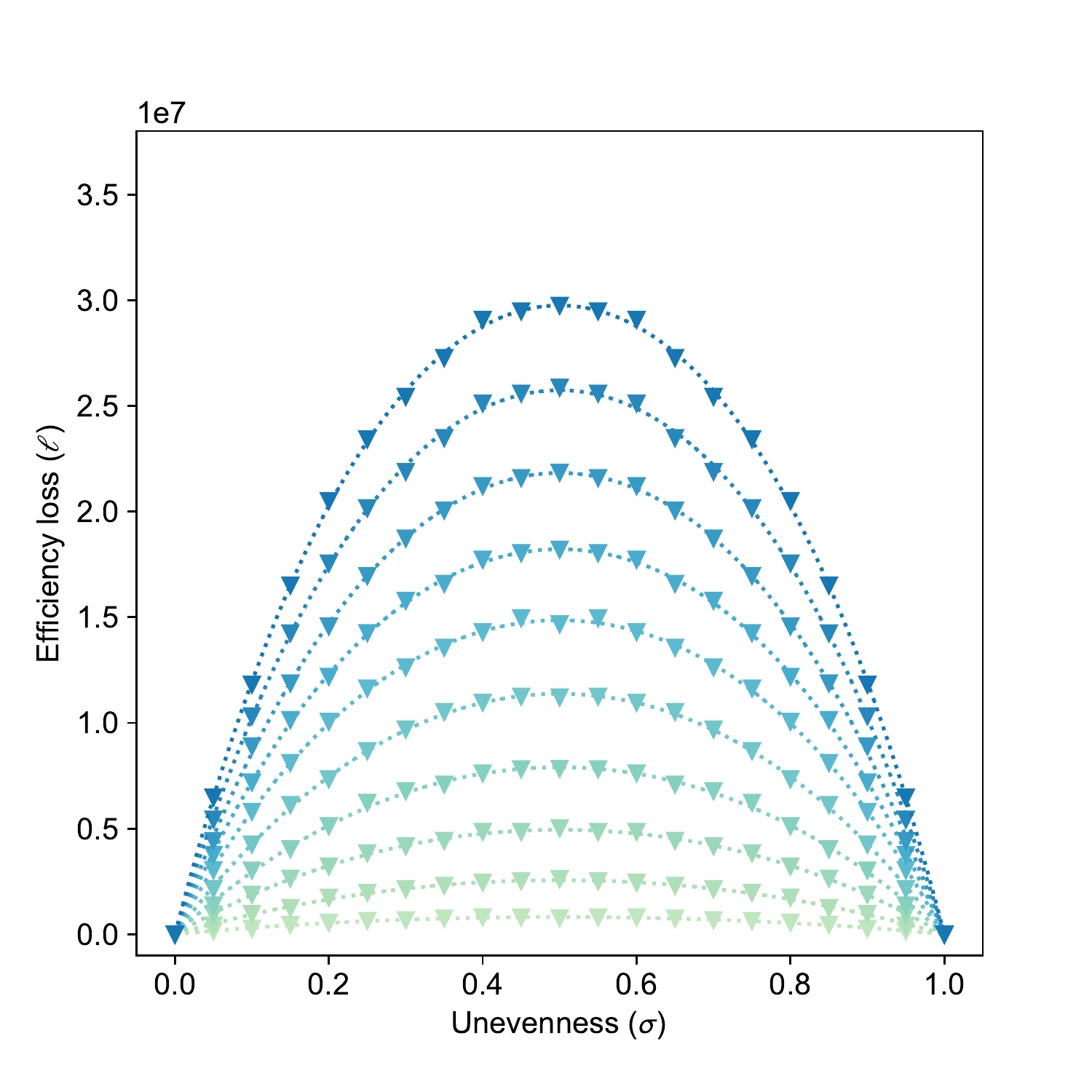}
        \put(-150,145){\footnotesize \fontfamily{phv}\selectfont (a)}
    \put(-130,135){\tiny\fontfamily{phv}\selectfont $\delta$ = 30 s}
    \includegraphics[height=2.1in,trim=0.7in 0.2in 0.55in 0.5in, clip]{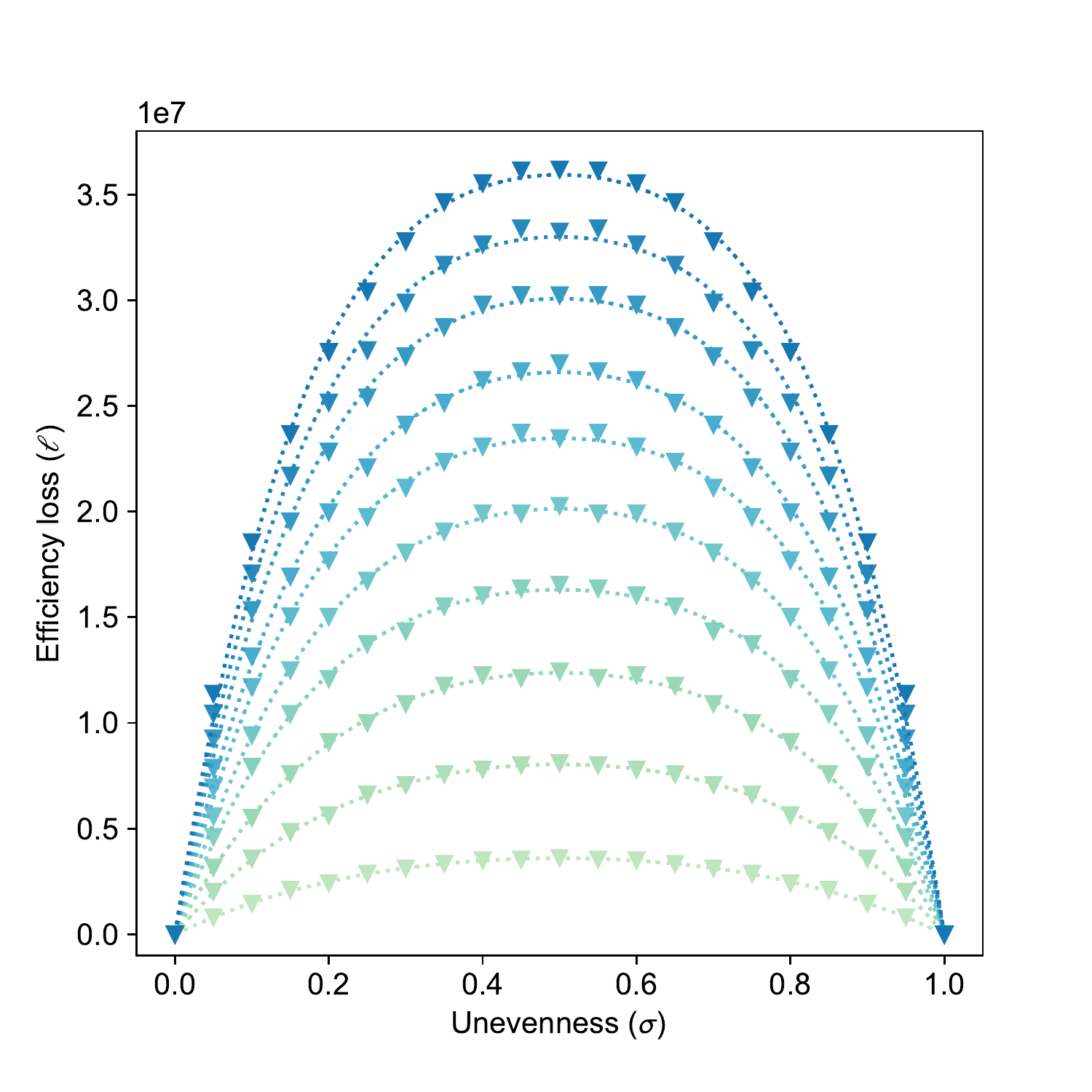}
    \put(-130,135){\tiny\fontfamily{phv}\selectfont $\delta$ = 120 s}
    \includegraphics[height=2.1in,trim=0.7in 0.2in 0.55in 0.5in, clip]{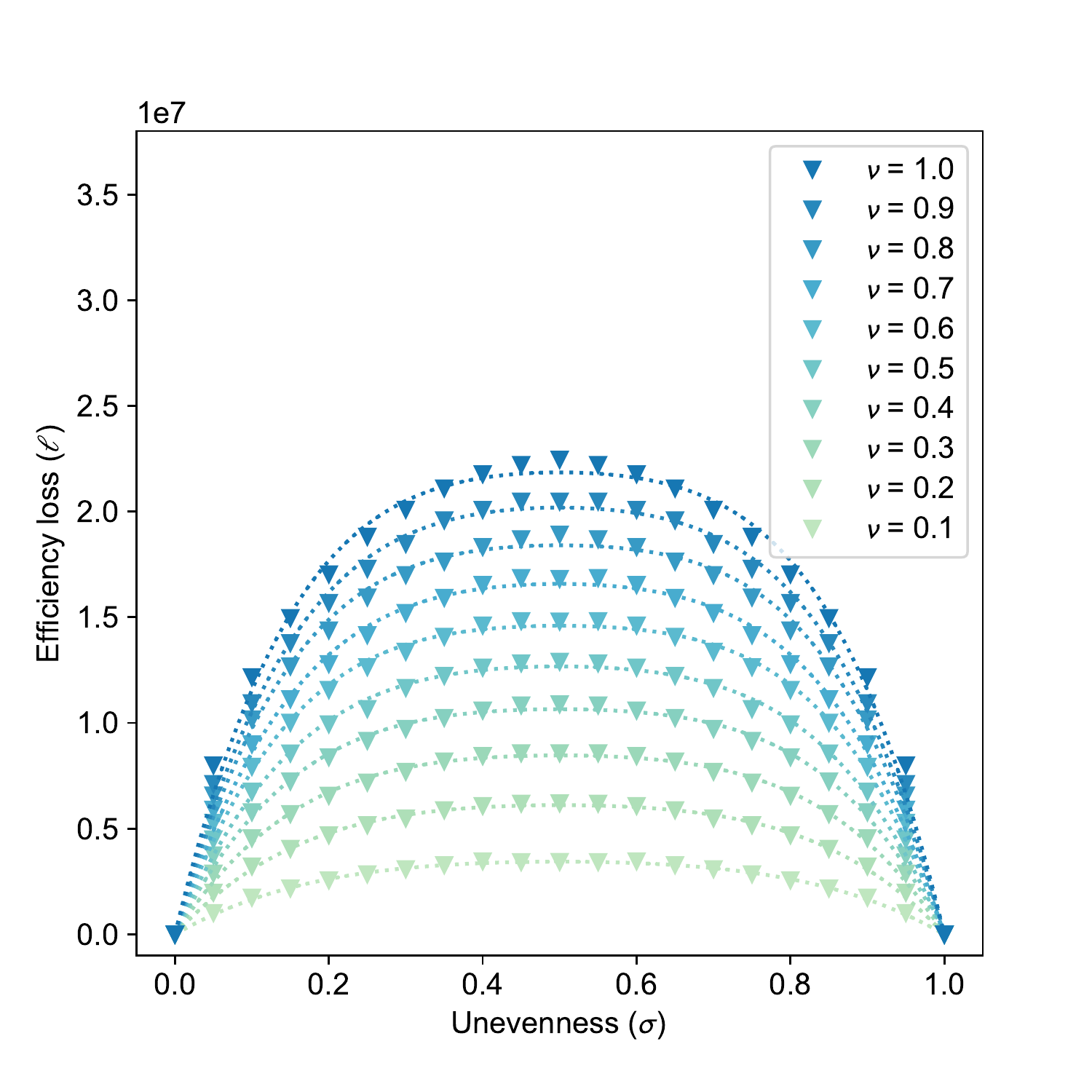}
    \put(-130,135){\tiny\fontfamily{phv}\selectfont $\delta$ = 300 s}
    
    \includegraphics[height=2in, trim=2.7in 1.3in 6in 1.4in, clip]{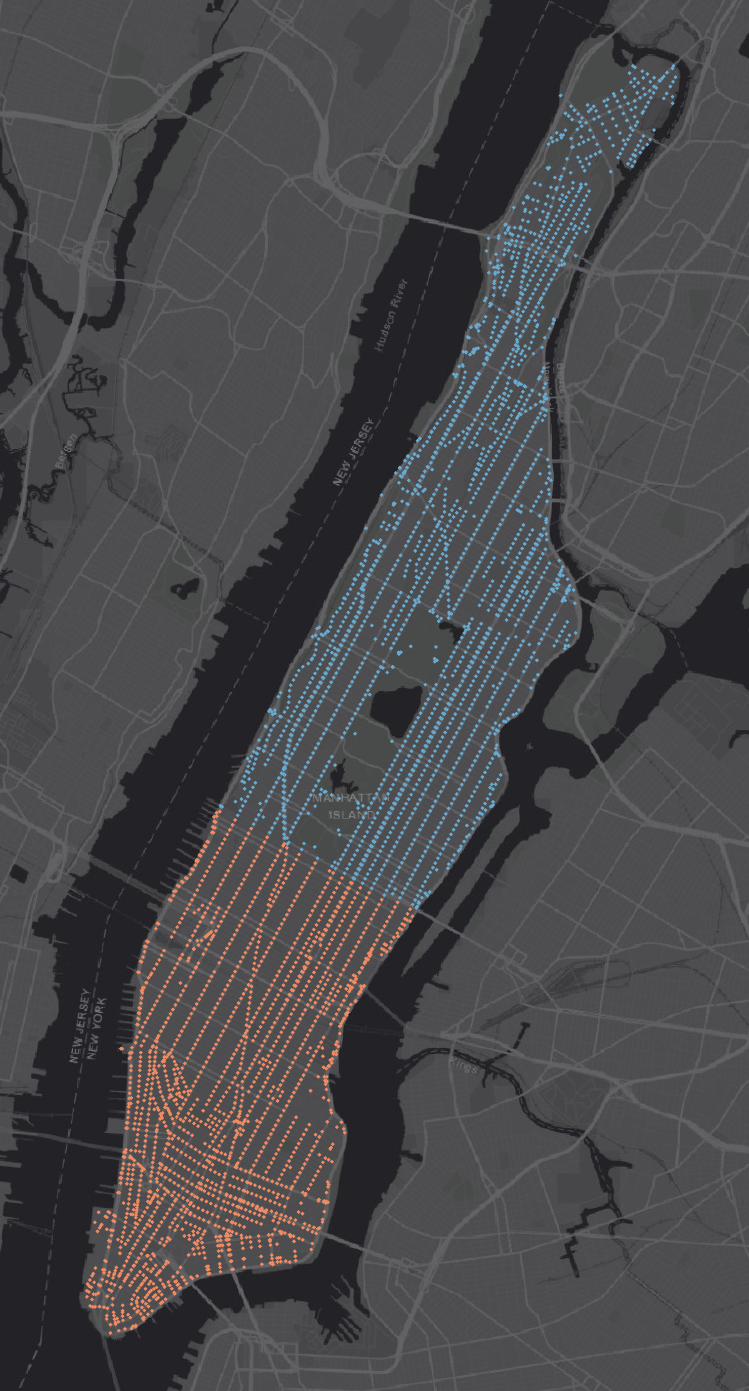}
    \put(-75,132){\footnotesize \fontfamily{phv}\selectfont \color{white}(b)}
    \includegraphics[width=3in,trim=0in 0.02in 0.55in 0.2in, clip]{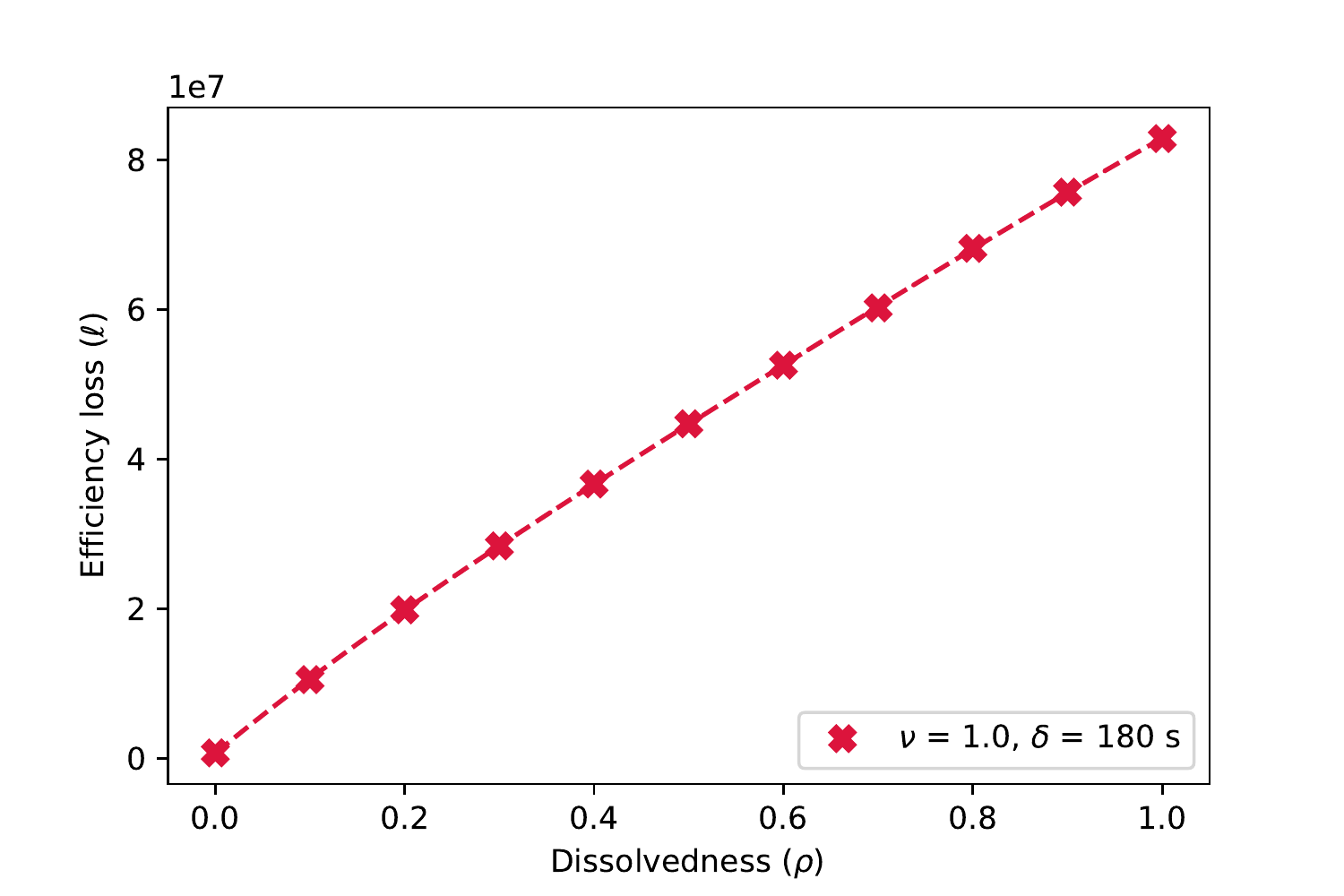}
    \put(-210,132){\footnotesize \fontfamily{phv}\selectfont (c)}
    \caption{Efficiency loss with unevenness (fitted with quartic functions) and dissolvedness}
    \label{fig:uneven-dissolve}
\end{figure}

The relationship between efficiency loss and unevenness, or the market shares of the two TNCs, shows a simpler pattern, since even intuitively we may guess that the efficiency loss will be the largest when the two market shares are half and half. However, we are still interested in the explicit form. Fixing any $\delta$ and $\nu$ we are able to depict the curves, as shown in Fig. \ref{fig:uneven-dissolve}a. The best fitted curves for all $\delta$'s and $\nu$'s are the quartic functions (Eq. \ref{eq:unevenness}). The derivation of the fitted form can be found in the Supplementary Information. Comparing the heights of the corresponding curves in the three panels, again as we have observed in Fig. \ref{fig:mainresults}f, they rise and fall with $\delta$.

\begin{equation}
    \ell(\sigma) = a\sigma^4 -2a\sigma^3 + c\sigma^2 + (a-c)\sigma.
    \label{eq:unevenness}
\end{equation}

The dissolvedness parameter $\rho$ describes the spatial segregation of the duopolistic TNCs, denoted as if each TNC has a base zone (Uptown vs. Mid- and Downtown in Manhattan), the proportion of trips which can be randomly drawn from \emph{the other} zone---being zero means that the two are completely segregated, such as the case in Fig. \ref{fig:uneven-dissolve}b, and being non-zero means that each platform can still take a proportion of trips from the other zone. The result in Fig. \ref{fig:uneven-dissolve}c shows that the higher the spatial segregation is (the left-hand side), the lower the efficiency loss will be. This is partially counter-intuitive, but actually explainable: Spatially proximate trips are more likely to be matched. Therefore, if the market is segmented based on spatial proximity, there will be smaller efficiency loss. Actually, this is the case for the taxi markets in many cities, such as the yellow cabs in Manhattan and green cabs in the outer boroughs of New York City.

\section{Discussion}
To summarize, we have the following findings in the paper:
\begin{enumerate}
    \item The two forces driving the economies and diseconomies of scale in ridesharing---network effect vs.\,Wild Goose Chase (WGC, indicated by the increasing average detour time with $\delta$) \cite{Castillo2017,Xu2019}: an analogy to the intermolecular energy potential. The analogy is both morphological in terms of the shape of curves, formal in terms of the fitted formulae, and explainable in terms of the decomposition of an attractive force and a repulsive force.
    \item Both the relative strengths and the positions of turning points of the two forces differ in the monopoly market and in the segmented market, leading to the non-linear and non-monotonic relationship of efficiency loss with thickness (the density of graph nodes) and tightness (the density of graph edges).
    \item Market share also impacts the efficiency loss, with the maximum loss at 50--50 market shares.
    \item Localized market segmentation can mitigate the efficiency loss compared with spatially random segmentation.
\end{enumerate}

We would like to expand the discussion to more general cases in several aspects. The fundamental reason of the non-monotonicity is the increasing marginal cost of matching, which has a spatial connotation---the pick-up distance. In the markets where there are higher spatial frictions, such as transportation, lodging, or delivery, the expansion of market will inevitably see higher marginal cost when the service area enlarges. Therefore, a spatial segmentation may reduce the inefficiency because it ``contains'' this spatial expansion.

The inefficiency, as represented by VMT in this paper, can be further translated to carbon emissions and energy consumed. Therefore, the peaked-shape curves indicate that although we know the ``optimal'' market in terms of both total VMT and efficiency loss only in the infinite horizon of $\delta$---the longer flexibility people would allow, the more energy saving there will be---the ``worst'' $\delta$ is well in the finite domain. For each given market of certain parameters of trip densities, market shares, spatial segregatedness, etc., we are able to calculate the worst sharing flexibility. We denote the frontier of the ``worst'' cases as the ``weakest market structures'' (e.g., the line connecting the hexagons in Fig. \ref{fig:mainresults}f) and give a full description of them in the Supplementary Information.

Moreover, a further question to ask is how we may be able to reduce the inefficiency of market segmentation while keeping the effect of healthy competition. We noticed that the primary reason of inefficiency in a segmented market is because it ``cuts'' the links of potential matching, which could have existed in a monopolistic market. Therefore, the direction would be towards fixing those missing links. A market design of allowing cross-platform matching would be viable options, such as adding a central coordinator, a trip broker, or by allowing information sharing or mutual trip trading. By adding back the missing links, the efficiency should be moving closer to the monopoly case.

The similarity between the inefficiency curves in this paper and the Lennard--Jones model is rhetoric, formal, and also substantive. The trips in a mobility sharing market are comparable to molecules---while they may appear like \emph{static} nodes in a given shareability network, the tightness of matching  changes their relative distances in the market. This parallel example of similar phenomena in natural and social sciences has also been discovered for other cases, including the segregation model \cite{vinkovic2006physical}.  We hope this analogy would not only be explanatory in the economics of sharing markets, but also informative for the natural sciences.

\section*{Methods}
\subsection*{Dataset.} 

The taxi data of Manhattan, New York on April 24, 2011 containing 301,430 trips are used in this paper. The dataset includes the following trip information: medallion ID, pick-up time, drop-off time, pick-up location, and drop-off location. The Street network of Manhattan is constructed with data from OpenStreetMaps and each trip's origin and destination are matched to the nearest nodes (intersections) in the street network with a distance shorter than 100 meters. Trip lengths are estimated with the shortest-path distances from origins to destinations in the constructed street network. Very short trips ($<$ 100 meters) are excluded from the analysis.

\subsection*{Proposed VMT Calculation.} 

To quantify VMT$_0$, VMT$_1$ and VMT$_2$ proposed in this paper, we first establish a complete \emph{shareability network} based on the trip data. The \emph{shareability network} proposed by \citet{Santi2014} is an undirected graph representing sharing potentials between trips. In the shareability network, each node indicates a trip and an edge exists between two nodes if they can potentially be shared and connected under necessary time constraints. The time constraints in the shareability network are represented by $\Delta$, indicating the largest \emph{delay} of passengers in a shared trip. The \emph{delay} time for a passenger in a shared trip is the time difference between the shared trip and the dedicated trip. The maximum allowed delay time is set to be 10 minutes when constructing the shareability network.

After having a complete shareability network $G$ containing all trips, we sample a sub-graph based on the thickness parameter $\nu$. The sampling procedure keeps the $\nu$ proportion of nodes and edges between them in the shareability network. The tightness parameter $\delta$ represents the maximum allowed delay time for shared trips. Therefore, only edges with a delay time less than $\delta$ seconds are kept in the sub-graph. The first two steps lead to a shareability network $\Tilde{G}(\nu, \delta)$ under a scenario with a monopoly TNC. To construct shareability networks for two TNCs in a duopoly market, we sample the shareability network $\Tilde{G}_1(\nu, \delta, \rho, \sigma)$ for one platform according to dissolvedness $\rho$ and unevenness $\sigma$. For a TNC with $\sigma$ market share, i.e., $\sigma$ proportion of nodes in the shareability network $\Tilde{G}(\nu, \delta)$, $\rho$ proportion of its nodes will be sampled uniformly at random while the remainder is sampled based on a fixed spatial distribution shown in Fig. \ref{fig:uneven-dissolve}b. Given the sampled nodes for one platform, the shareability network of the other platform $\Tilde{G}_2(\nu, \delta, \rho, \sigma)$ can then be generated by excluding nodes for the first platform in the sub-graph.

With generated shareability networks, edge weights indicating the total VMT savings due to sharing are calculated. The maximum weight matching algorithm implemented in the NetworkX, which is based on methods proposed by \citet{Galil1986}, is utilized in this paper. The maximum weight matching algorithm produces the maximum VMT savings due to sharing given a shareability network. The maximum weight matching can be solved in polynomial time $O(n^3)$ where $n$ represents the number of nodes in a shareability network.

Given a set of parameters $\nu, \delta, \rho$ and $\sigma$, shareability networks $\Tilde{G}(\nu, \delta)$, $\Tilde{G}_1(\nu, \delta, \rho, \sigma)$ and $\Tilde{G}_2(\nu, \delta, \rho, \sigma)$ are generated. VMT$_0$ is the summation of trip distance of nodes from $\Tilde{G}(\nu, \delta)$. VMT$_1$ is calculated by subtracting the total VMT saving (calculated with the maximum weight matching algorithm) of $\Tilde{G}(\nu, \delta)$ from VMT$_0$. VMT$_2$ is calculated by subtracting the total VMT savings of $\Tilde{G}_1(\nu, \delta, \rho, \sigma)$ and $\Tilde{G}_2(\nu, \delta, \rho, \sigma)$ from VMT$_0$.

\bibliographystyle{unsrtnat}
\bibliography{segmentation}



\end{document}